# Potassium polytungstate nanoparticles by combustion aerosol technology for benzene sensing

Adrien Baut, Sebastian Kravecz and Andreas T. Güntner[*]

Human-centered Sensing Laboratory, Department of Mechanical and Process Engineering, ETH Zurich, CH-8092 Zurich, Switzerland.

[*] Corresponding author: andregue@ethz.ch

# Abstract

Polytungstates are oxygen-linked assemblies of highly oxidized tungsten polyhedra, valued for their tunability and stability in diverse applications. Traditional synthesis methods (hydrothermal, solvothermal, solid-state) offer material variety but are limited in scalability and their ability to yield nanostructured materials due to long reaction times and high temperatures. Here, we introduce flame aerosol synthesis as a single-step, rapid and dry method to prepare $K_2W_7O_{22}$ nanoparticulate powders and coatings. Thereby, monocrystalline and phase-pure $K_2W_7O_{22}$ with varying crystal-sizes were obtained by controlling flame temperature, residence time and metal ion concentration during particle formation by nucleation, coagulation and sintering. X-ray diffraction and electron microscopy identified the high potassium tolerance of the $K_2W_7O_{22}$ lattice (K/W ratio up to 0.6) and phase stability up to 400 °C, before other polytungstates and $WO_3$ polymorphs were formed, respectively. Porous films of such $K_2W_7O_{22}$ nanoparticles featured n-type semiconductor behavior that was utilized for the chemoresistive quantification of the air pollutant benzene down to 0.2 parts-per-million at 20% relative humidity. Such sensors were quite selective over other compounds (e.g. alcohols, aldehydes, ketones, CO, $NH_3$ or $H_2$), in particular to chemically similar toluene and xylene (>18).

# Introduction

Polytungstates are complex structures usually formed by edge and corner sharing $WO_x$ (x = 3 - 7) polyhedral units[1,2] with the W atoms in a high oxidation state. They have been widely used in the form of complex polyoxotungstates (clusters of polyatomic ions[3]) as antiviral[4], antibacterial[5] and anticancer[6] materials. Polytungstates can also be found in well-defined solid structures when the $WO_x$ polyhedra are arranged into regular structures. In particular, when $WO_6$ octahedra are present, hexagonal channels can be formed able to accommodate alkali[7] (e.g. K, Cs, Rb) or other (e.g. Tl[8]) elements to form slabs of various sizes. Such structured materials have shown practical applications in the design of lasers[9], catalysts[10] or sensors[11-13]. For instance, $K_2W_4O_{13}$ has been used to detect $H_2S$ down to 300 parts-per-billion[11], $Rb_4W_{11}O_{35}$ was able to sense changes in relative humidity within 1 s[12], while $K_2W_7O_{22}$ showed promising results for the detection of acetone in high humidity environments[13].

Traditionally, polytungstates are synthesized by hydrothermal, solvothermal or by solid-state reaction. The first two methods rely on the condensation of $WO_x$ (x = 3-7) building units[1] where synthesis parameters like acidity of the medium[14], temperature[14] and duration[14] all influence the structure of the final product and its size (e.g. from 20-30 nm $K_2W_4O_{13}$ nanoparticles for a hydrothermal reaction of 30 min to a few μm for a 2 h one[11]). These procedures include multiple synthesis and cleaning steps, resulting in rather long times (e.g. > 15 h[11] including all the steps) to obtain the desired purified product. Alternatively, solid-state reactions between tungstate containing species (e.g. $K_2WO_4$[7]) and $WO_3$ in air at high temperatures (e.g. 500 - 900 °C[7]) and for prolonged times (e.g. 3-5 days[7]) also yield polytungstates, with sizes sometimes in the μm[7] scale due to sintering. The scalability of these traditional syntheses to produce nanoscaled particles is therefore limited due to the long production times and/or high reaction temperatures needed.

Here, we pioneer the use of rapid combustion-aerosol synthesis to self-assemble nanoscaled potassium polytungstate $K_2W_7O_{22}$ by gas-solid conversion. Flame spray pyrolysis (FSP) is a one-step dry nanoparticle[15] and thin film[16] production technique that has proven scalability (up to kg/h[17]). The crystal phase composition, crystal and particle size of the obtained powder as a function of the K to W atomic ratio was assessed by X-ray diffraction (XRD) and nitrogen desorption analysis. High resolution transmission electron microscopy (HRTEM) and elemental mapping characterized the particles and homogeneity of atom distribution. The influence of synthesis parameters on material composition and morphology was investigated. Also, the thermal stability of $K_2W_7O_{22}$ up to 800 °C in air and their application for chemoresistive sensing were evaluated to demonstrate immediate practical impact.

## Methods

*Nanoparticle powder synthesis*

The particles were produced via FSP, using a reactor detailed elsewhere[18]. The precursor solutions contained 0.2 M of K and W, as dictated by the desired K/W ratio. Therefore, two separate solutions were prepared and only mixed no longer than 2 min before the FSP, as some precipitate formation had been observed after approximately 6 min. The first solution consisted of ammonium metatungstate hydrate (> 85%, Merck, Switzerland) dissolved in diethylene glycol monobutyl ether (Merck, Switzerland) under vigorous stirring overnight at room temperature. The second solution was potassium carbonate (Merck, Switzerland) dissolved in methanol (Merck, Switzerland) by magnetically stirring the solution for a few minutes. The precursor was fed through a capillary at rate P (in ml/min) and dispersed by oxygen (technical grade, PanGas, Switzerland) with rate D (in l/min). The P/D ratio was set to 5/5, unless specified differently in the text. A pressure drop for the dispersion gas was maintained at 1.6 bar. A circular pilot flame of premixed methane (1.25 l/min, Methane 2.5,

PanGas, Switzerland) and oxygen (3.25 l/min, technical grade, PanGas, Switzerland) ignited and sustained the spray flame. A sheath gas of oxygen (5 l/min, technical grade, PanGas, Switzerland) surrounding the premix flame was supplied to protect the flame and ensure an oxygen-rich atmosphere. The particles were captured on a water-cooled glass microfiber filter (GF-6 Albert-Hahnemuehle, 257 mm diameter) positioned 57 cm above the nozzle and assisted by a vacuum pump. The powder was collected by scraping it from the filter using a spatula before passing it through a 250 μm mesh sieve to eliminate any filter residues.

*Film deposition*

Porous films were thermophoretically deposited[19] onto $Al_2O_3$ substrates (electrode type #103, Electronic Design Center, Case Western University, USA), featuring interdigitated Pt electrodes on the front and a back Pt heater element. The substrates were fixed onto a water-cooled holder and positioned 20 cm above the FSP reactor. The deposition time was 4 min. To ensure a good adhesion to the substrate, the films were then annealed for 5 h at 250 °C in an oven (CWF 1300, Carbolite Gero, Germany) under atmospheric condition.

*Powder characterization*

The X-ray diffraction (XRD) patterns were obtained using a Bruker D2 Phaser (Bruker, USA) operated with Cu $K_\alpha$ radiation at 30 kV and 10 mA. The scanning was done for angles $2\theta$ of 10 – 65°. with a scanning speed of 1.1 s/step and a scanning step size of 0.012°. The identification of the crystal phase was done with the Diffrac.eva V3 software and reference patterns of ε-$WO_3$ (PDF 24-0747), γ-$WO_3$ (PDF 83-0950), β-$WO_3$ (PDF 71-0131), $K_2W_7O_{22}$ (PDF 21-0700), $K_2W_8O_{25}$ (PDF 31-1114), $K_2CO_3 \cdot 1.5\ H_2O$ (PDF 11-0655) and $K_4H_2(CO_3)_3 \cdot 1.5\ H_2O$ (PDF 20-0886). The crystal sizes and composition of powders were determined with the Rietveld refinement method on the TOPAS 4.2 software between $2\theta = 20 – 65°$. XRD patterns at elevated temperatures were collected *in situ* in a high-temperature non-ambient chamber (HTK 1200N, Anton Paar, Austria) mounted onto a Bruker AXS D8 Advance diffractometer operated at 40 kV and 40 mA for angles 2θ of 10-65°. The applied temperature profile is shown in Figure S1.

The specific surface area (SSA) was measured via nitrogen adsorption at 77 K (Tristar II Plus, Micromeritics, USA) with the Brunauer-Emmett-Teller (BET) 8-points method. The powders were freed from surface adsorbates by heating them to 120 °C for 1 h 15 min in a $N_2$ atmosphere. The BET equivalent diameters ($d_{BET}$) were calculated as:

$$d_{\mathrm{BET}} = \frac{6000}{\rho \cdot SSA}$$

The density ρ was estimated as follows:

$$\rho = \frac{1}{\sum_i \frac{w_i}{\rho_i}}$$

where the weight fractions $w_i$ were the amount of $WO_3$ and $K_2W_7O_{22}$ determined from the XRD patterns, while the densities $\rho_i$ were 7.16 and 6.42 g/cm³ for $WO_3$ and $K_2W_7O_{22}$[7], respectively.

HRTEM particle images were obtained on a Grand-ARM300F (JOEL, Japan), equipped with a cold field emission gun operated at 300 kV. The aberration corrections enable atomic-resolution imaging in TEM mode. A few droplets of the powders dispersed in ethanol were deposited onto a perforated carbon foil supported on a molybdenum grid to perform the analysis. Afterwards, the grid was mounted on the single tilt holder of the microscope.

*Gas sensing characterization*

For the gas sensing performance evaluation, the deposited films were mounted onto Macor holder and loaded inside a PTFE chamber. The gas mixtures were controlled using a setup described previously[20]. The resistance of the sensing films was monitored continuously during the gas exposure using a multimeter (Keithley, DMM 7510 1, USA). The temperature of the sensor was adjusted by applying a constant voltage through the back Pt heater. The tested gas mixtures were prepared with mass-flow controllers (Bronkorst, Netherlands) to adjust their composition. Therefore, dry synthetic air (PanGas, 5.0, $C_nH_m$ and $NO_x \leq 100$ ppb) was humidified by partially bubbling it through deionized water at room temperature. Analytes were admixed from certified gas standards (all PanGas, with synthetic air as balance): benzene (14.9 ppm), xylene (10.3 ppm), toluene (15 ppm), ethanol (15 ppm), CO (506.2 ppm), $NO_2$ (18.0 ppm), formaldehyde (20.3 ppm, in $N_2$), $NH_3$ (10 ppm), $H_2S$ (10.7 ppm), $H_2$ (50 ppm) and acetone (20 ppm). All the transfer lines were made of Teflon and heated at 55 °C to mitigate the water condensation and analyte adsorption. The response ($S$) was calculated as $\frac{R_{air}}{R_a} - 1$ for reducing analytes and as $\frac{R_a}{R_{air}} - 1$ for oxidizing analytes, where $R_{air}$ is the baseline resistance of the film in air and $R_a$ the steady-state resistance of the film when exposed to the analyte gas. The selectivity was defined as $S_{ben}/S_{conf}$, with $S_{ben}$ and $S_{conf}$ the response when the film is exposed to benzene and to confounding gas, respectively.

## Results and Discussion

*Dry synthesis of $K_2W_7O_{22}$ nanoparticles*

Figure 1a shows the powder XRD patterns for various potassium to tungsten atomic ratios (K/W). For pure tungsten (i.e. K/W = 0.0), the powder is highly crystalline, with peaks associated to γ- (stars) and the metastable ε-$WO_3$ (squares) phases, in agreement with literature[21]. When potassium is introduced, the XRD pattern gradually changes. In particular, a small hump at 2θ = 14° attributed to $K_2W_7O_{22}$ (diamonds) appears already at K/W = 0.05. Pure $K_2W_7O_{22}$ is obtained starting at K/W = 0.29 (see also Figure 2a, squares and right ordinate for amount), as expected from its stoichiometric composition and other studies[22].

During FSP, the liquid precursor is rapidly evaporated (< 1 ms for a different precursor composition[17]). Due to the high temperature, tungsten and potassium ions feature high mobility leading to rapid particle formation by nucleation and coagulation[15]. Apparently, $WO_6$ structures are formed by oxidation that quickly self-assemble to the final $K_2W_7O_{22}$ crystals composed of hexagonal channels of such edge and corner sharing octahedra. Due to the short residence time in the flame (few ms[17]), the crystals remain nanosized. As a result, pure $K_2W_7O_{22}$ powder is obtained in a single step and rapidly by FSP, bypassing more time-consuming hydrothermal synthesis[13] or solid-state reaction[7]. The production rate with our laboratory based FSP reactor is 3.4 g/h (at K/W = 0.29) and pictures of powder samples at various K/W ratio are shown in Figure 1b.

Interestingly, even when the stoichiometry is further increased up to K/W = 0.6, $K_2W_7O_{22}$ remains the only crystalline phase detected (Figure 1a and Figure 2a) suggesting high solubility of potassium without interference of the lattice structure. Previous studies on the solid-state reaction of $K_2WO_4$ and $WO_3$ at 900 °C had also reported the presence of potassium-rich (up to K/W ≈ 0.6) $K_2W_7O_{22}$, although some liquid phase was co-present[7]. When increasing K/W to 0.8 and 1 (Figure 1a), the peaks of $K_2W_7O_{22}$ vanish and only broader humps with peaks at 2θ = 11.6° and 28.2° are visible, suggesting an amorphization of the particles. Finally, for pure potassium, potassium carbonate hydrate species (triangles and circles) are observed (Figure 1a). Their presence is likely due to the reaction of $K_2CO_3$ with water vapor[23] formed during the combustion.

Figure 2a shows the BET-equivalent particle size ($d_{BET}$, green circles) and the crystal size of the $K_2W_7O_{22}$ phase ($d_{XRD}$, blue triangles). Pure $WO_3$ features average particle size of 13.1 nm that is comparable to HRTEM (Figure 2b) and to literature[21]. The $WO_3$ particles are monocrystalline, as observed on a magnified particle (inset, Figure 2b) where the lattice fringes of the $(\bar{1}10)$ plane of ε-$WO_3$ extend over the entire particle. When increasing potassium content to the stoichiometric ratio of $K_2W_7O_{22}$ (i.e. K/W = 0.29), the particle size decreases to 8.9 nm, in line with the rather similar crystal size of 8.3 nm. The crystal structure and monocrystallinity of $K_2W_7O_{22}$ are further confirmed by HRTEM (Figure 2c), where lattice fringes associated to the $(20\bar{1}0)$ plane are identified. At higher potassium content (e.g. K/W = 0.5), the nanoparticles become partially amorphous (Figure S2) until no lattice structure is visible anymore at K/W = 1 (Figure 2d), confirming the earlier observations in the XRD patterns (Figure 1a).

*Process parameters effect on synthesis of $K_2W_7O_{22}$*

By modifying P/D, the length of the flame and its temperature profile above the burner as well as the metal ion concentration in the flame are tuned, as shown in previous work by computational and experimental methods[17,24,25]. In particular, increasing the value of D, results in shorter flames (pictures in Fig. 3b) and smaller particle residence times in the flame[17]. As a consequence, the product aerosols are diluted and their growth hindered as sintering effects and coagulation are reduced. Furthermore, reducing the value of P leads to shorter and lower enthalpy flames, i.e. colder flames, with a decreased ion concentration[25]. Modifying P/D thus leads to control over the particle size and phase composition, with even access to metastable materials (e.g. $CoCu_2O_3$[26], $\varepsilon$-$WO_3$[21] or $In_4Sn_3O_{12}$[27]) due to the rapid quenching of the flame. Interestingly, only P/D values between 4/6 up to 6/4 at K/W = 0.29 yielded pure $K_2W_7O_{22}$ (Figure 3a). For P/D = 2/8 and 3/7, additional phases of $\gamma$-$WO_3$ and polytungstate $K_2W_8O_{25}$ were identified. The transition from $K_2W_7O_{22}$ to $K_2W_8O_{25}$ is still unclear in literature, as early reports indicate a continuous evolution by modifying K/W ratio for solid state synthesis[22], that was not confirmed in more recent work[7].

The specific surface area (SSA) and $K_2W_7O_{22}$ crystal size as a function of P/D ratio are shown in Figure 3b (insets show corresponding FSP flames). Between P/D = 2/8 – 4/6, the SSA increases from 60 to 110 $m^2/g$ when pure $K_2W_7O_{22}$ is obtained (Figure 3a). With higher P/D ratio, the SSA decreases to 81 $m^2/g$, as expected due to the longer particle residence time at high temperature increasing sintering and coagulation. This is also reflected by the increasing crystal size of $K_2W_7O_{22}$ from 7.6 to 9.6 nm between P/D = 4/6 and 6/4, respectively. Please note that $d_{XRD}$ was not calculated for P/D = 2/8 and 3/7 due to the lack of crystallographic data for $K_2W_8O_{25}$ and overlapping peaks of the identified phases, rendering the Rietveld refinement impossible.

*Thermal stability of $K_2W_7O_{22}$*

To study the thermal stability of $K_2W_7O_{22}$, we performed *in situ* XRD while heating powder samples at K/W = 0.29 (Figure 4a) until 800 °C in air (see Figure S1 for the temperature profile). Until 400 °C, no change is identified, suggesting the preservation of $K_2W_7O_{22}$ (diamonds) crystals. Only at 500 °C, $K_2W_7O_{22}$ demixes and $WO_3$-related crystals are formed. In fact, a new peak around $2\theta = 42°$ emerges (vertical dotted line) while additional peaks between $2\theta = 23\text{-}25°$ and $33\text{-}35°$ become visible at 600 °C that are all characteristic for γ-$WO_3$. Increasing the temperature further to 800 °C, the high-temperature β-phase of polymorphic $WO_3$ (hexagons) is identified, as seen with the disappearance of the γ-$WO_3$ peak around $2\theta = 35°$ (vertical dashed line) and the change in relative peak intensities. The temperatures at which the γ- and β-$WO_3$ polymorphs appear are similar to earlier results obtained during the annealing of hexagonal-$WO_3$[28] (h-$WO_3$) to 900 °C in air. Other studies had reported the formation of β-$WO_3$ already at 370 °C[29] when heating room temperature stable γ-$WO_3$[29]. The higher temperature stability of the γ-$WO_3$ in our case (up to 700 °C) might be associated to the more stable structural arrangement of edge and corner sharing octahedra that both $K_2W_7O_{22}$ and h-$WO_3$[28] share. Remarkably, $K_2W_7O_{22}$ is present throughout the entire process (c.f. the peak around $2\theta = 14°$ in Figure 4a, vertical dash-dotted line), which coincides with the phase diagram for the $K_2WO_4$-$WO_3$ system[7]. This suggests that the K ions stabilize further the $K_2W_7O_{22}$ structure compared to h-$WO_3$ that completely transforms into other $WO_3$ polymorphs[28]. After cooling down the sample to 25 °C (top curve in Figure 4a), only peaks from γ-$WO_3$ and $K_2W_7O_{22}$ remain.

We also investigated the atomic dispersion of potassium (light green) and tungsten (red) before and after this annealing (Figures 4b,c and Figures S3a,b, respectively). Before annealing, we observed a homogeneous distribution of potassium and tungsten, in agreement with XRD (Figures 4a) where no other crystal phases had been detected. After the annealing (Figure 4c), two regions can clearly be seen – one composed of a homogeneous distribution of W and K (top part of the image) and the other one consisting mostly of W atoms (bottom of the image). This supports the co-presence $K_2W_7O_{22}$ and $\gamma$-$WO_3$.

*Chemoresistive properties of porous $K_2W_7O_{22}$ film*

To demonstrate immediate practical impact, the $K_2W_7O_{22}$ nanocrystals (K/W = 0.29) were deposited directly from the combustion-aerosol onto interdigitated electrodes on a water-cooled substrate via thermophoresis[19]. The obtained film is highly porous and formed by a fine network of interconnected fractal-like particles, as revealed from top-view scanning electron microscopy (SEM) images (Figures 5a-c). The film has a baseline around 600 M$\Omega$ at an operation temperature of 170 °C and under an air flow at 20 % relative humidity (RH). Upon exposure to various concentrations of benzene - a reducing analyte[30] - the resistance decreases, indicating an n-type semiconductor behavior[31].

Benzene is a highly toxic air pollutant, that the World Health Organization considers not even safe at ppb level[32] as prolonged exposure to benzene leads to higher risks in developing leukemia[33] and genetic alteration[34]. Remarkably, the $K_2W_7O_{22}$-based sensor can distinguish air quality-relevant benzene concentrations between 2.5 – 0.2 ppm with distinct resistance change (i.e. response, Figure 5e) and high signal quality. In fact, the signal-to-noise ratio is 42.6 at 0.2 parts-per-million (ppm, Figure 5d) with a theoretical (linearly extrapolated to a signal-to-noise ratio of 3) lower limit of detection of 0.014 ppm. Importantly, the sensor fully recovers after each benzene exposure (Figure 5d), indicating fully reversible analyte surface interaction. Note that the ability of the sensor to detect benzene is also preserved at higher relative humidity (RH), though at lower response (Figure S4).

Another important metric for an effective benzene sensor is its selectivity over other gases, particularly chemically similar aromatic compounds like xylene or toluene. Figure 5f shows the sensor response to a 1 ppm concentration of various classes of analytes that can be present in ambient air. The sensor responds mainly to benzene (response of 0.356) and only a little to the other compounds ($CO - H_2$), with selectivity values ranging from 5.6 – 29.7. In particular, the high selectivity to xylene and toluene (i.e. >18) are substantially higher than previous sensors also relying on metal oxide nanoparticles. For example, selectivities of only 5.0 and 1.66 towards toluene were reported for Pd-$TeO_2$[35] and Pd-rGO-ZnO[36] films, respectively. In case of high confounder concentrations, the selectivity can be further improved using filters[37,38] or surface clusters[39], that are compatible with handheld devices[40] and have already shown promising discriminating benzene from toluene and xylene[26,41].

## Conclusion

$K_2W_7O_{22}$ nanoparticulate powders were obtained in a single step by rapid and dry flame-aerosol synthesis. Such nanoparticles were monocrystalline and featured high potassium solubility tolerating K/W ratios of up to 0.6. The $K_2W_7O_{22}$ nanocrystals were stable in air until 400 °C, after which $WO_3$ polymorphs were partially formed. By varying the synthesis parameters, the temperature profile, particle residence time and metal ion concentration in the flame were modified, which resulted in particle size alteration and the formation of other polytungstates. Immediate practical impact was demonstrated by applying such $K_2W_7O_{22}$ nanoparticles as a porous film that showed n-type semiconductive behavior, and enabled the highly sensitive and selective chemoresistive detection of trace-level benzene vapors.

## Data availability

The data that support the findings of this study can be requested from the corresponding author.

## Acknowledgment

This study was financially supported by the Swiss State Secretariat for Education, Research and Innovation (SERI) under contract number MB22.00041 (ERC-STG-21 "HEALTHSENSE"). The authors also acknowledge Dr. Frank Krumeich for support with electron microscopy and the Scientific Center for Optical and Electron Microscopy (ScopeM) of ETH Zurich for providing measuring time on their instruments.

## Conflict of interest

The authors declare no conflict of interest.

## Author contributions

**Adrien Baut:** Conceptualization, Methodology, Investigation, Data Curation, Writing – Original Draft, Writing – Review & Editing, Visualization. **Sebastian Kravecz:** Methodology, Investigation, Data Curation. **Andreas T. Güntner**: Conceptualization, Methodology, Writing – Review & Editing, Visualization, Supervision, Funding Acquisition

## Additional information

**Supplementary information** is available for this paper.

**Correspondence and requests for materials** should be addressed to A.T.G.

# Figures and Tables

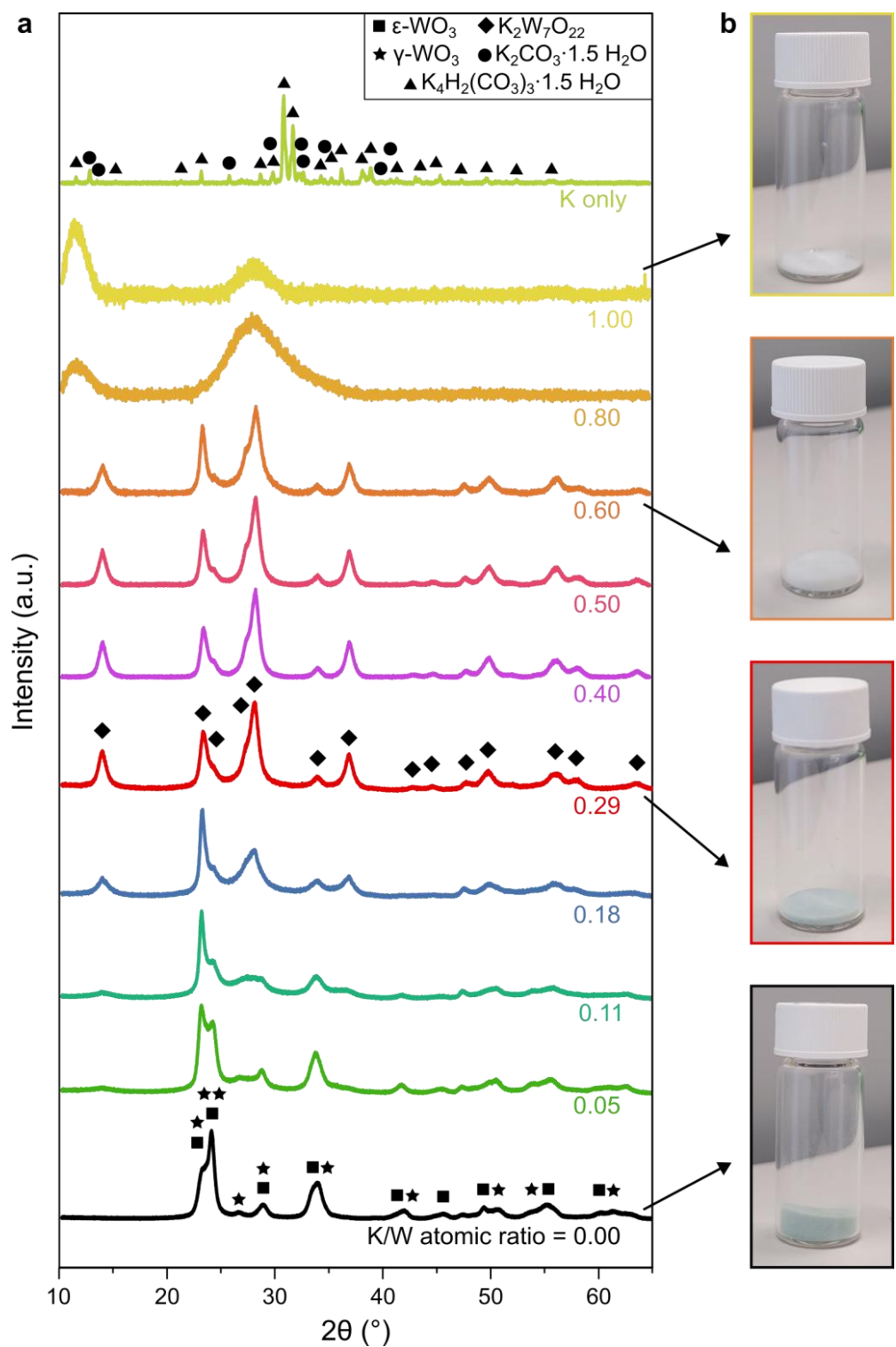

Fig. 1 (a) XRD patterns of as prepared potassium tungstate powder for different K/W atomic ratios. (b) Pictures of the collected powders.

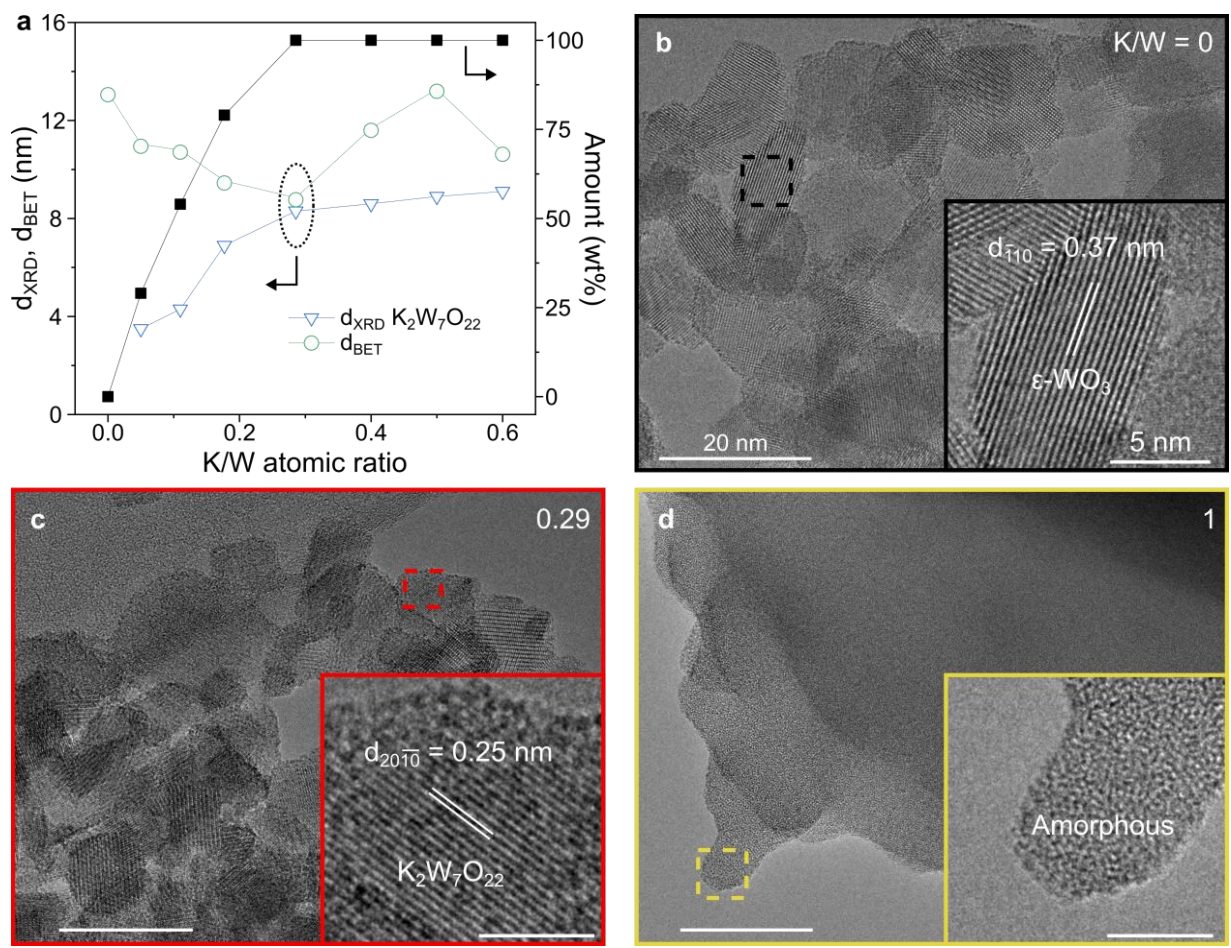

Fig. 2 (a) Particle (green circle) and crystal (blue triangle) sizes (both left ordinate) and weight fraction of $K_2W_7O_{22}$ (black square, right ordinate) as a function of the K/W atomic ratio. HRTEM images for (b) K/W = 0, (c) K/W = 0.29 and (d) K/W = 1.0. Insets show magnifications of the indicated areas in b-d.

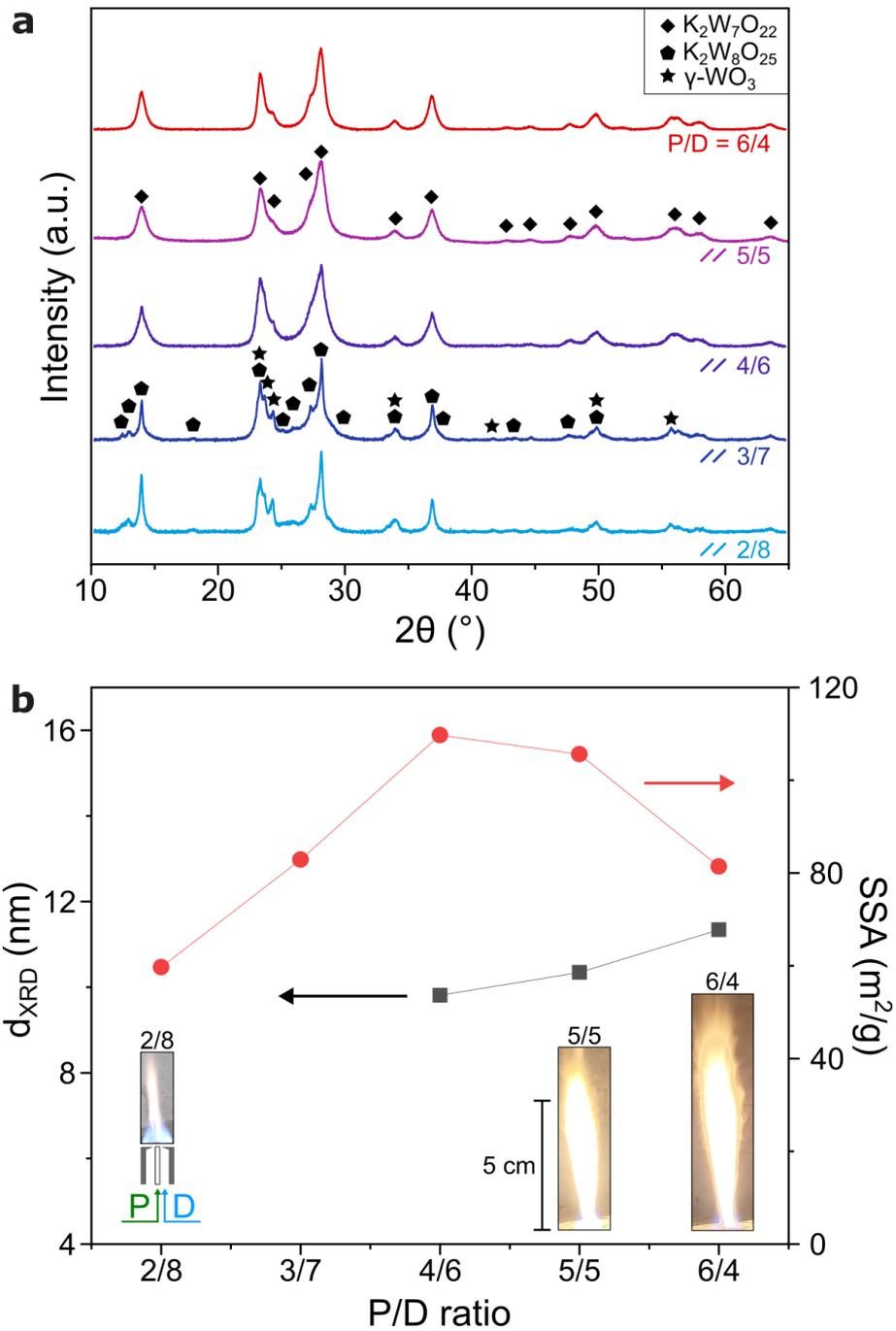

Fig. 3 (a) XRD patterns of $K_2W_7O_{22}$ particles for different values of P/D. (b) Crystal size (black square, left ordinate) and specific surface area (SSA, red circles, right ordinate) of the corresponding powders. Pictures of three flames of P/D ratio of 2/8, 5/5 and 6/4 are shown as insets.

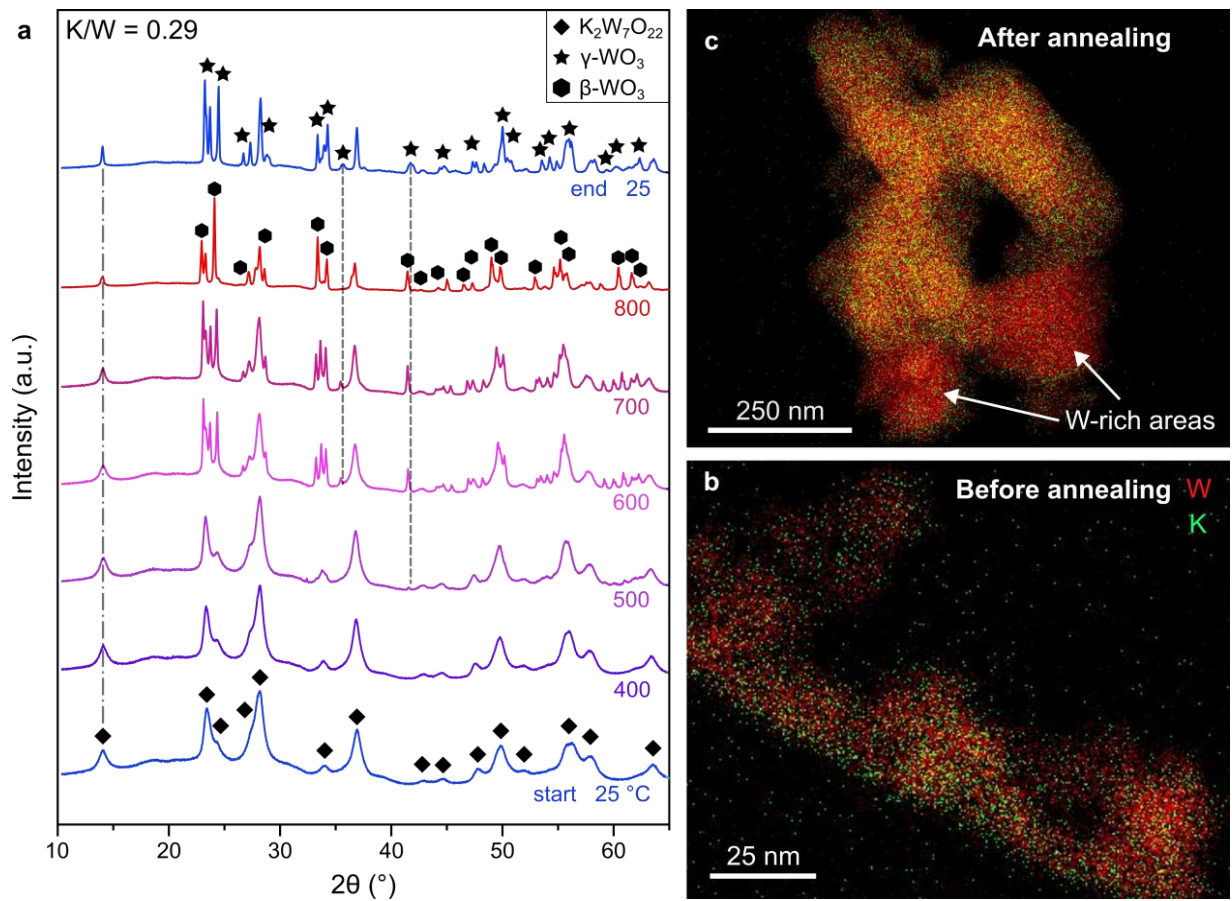

Fig. 4 (a) In situ XRD pattern of $K_2W_7O_{22}$ powders (K/W = 0.29) for annealing temperatures between room temperature and 800 °C for a total of 20.5 h in air. Corresponding elemental mapping of such powders (b) before and (c) after annealing. The W atoms are indicated in red and the K atoms in light green.

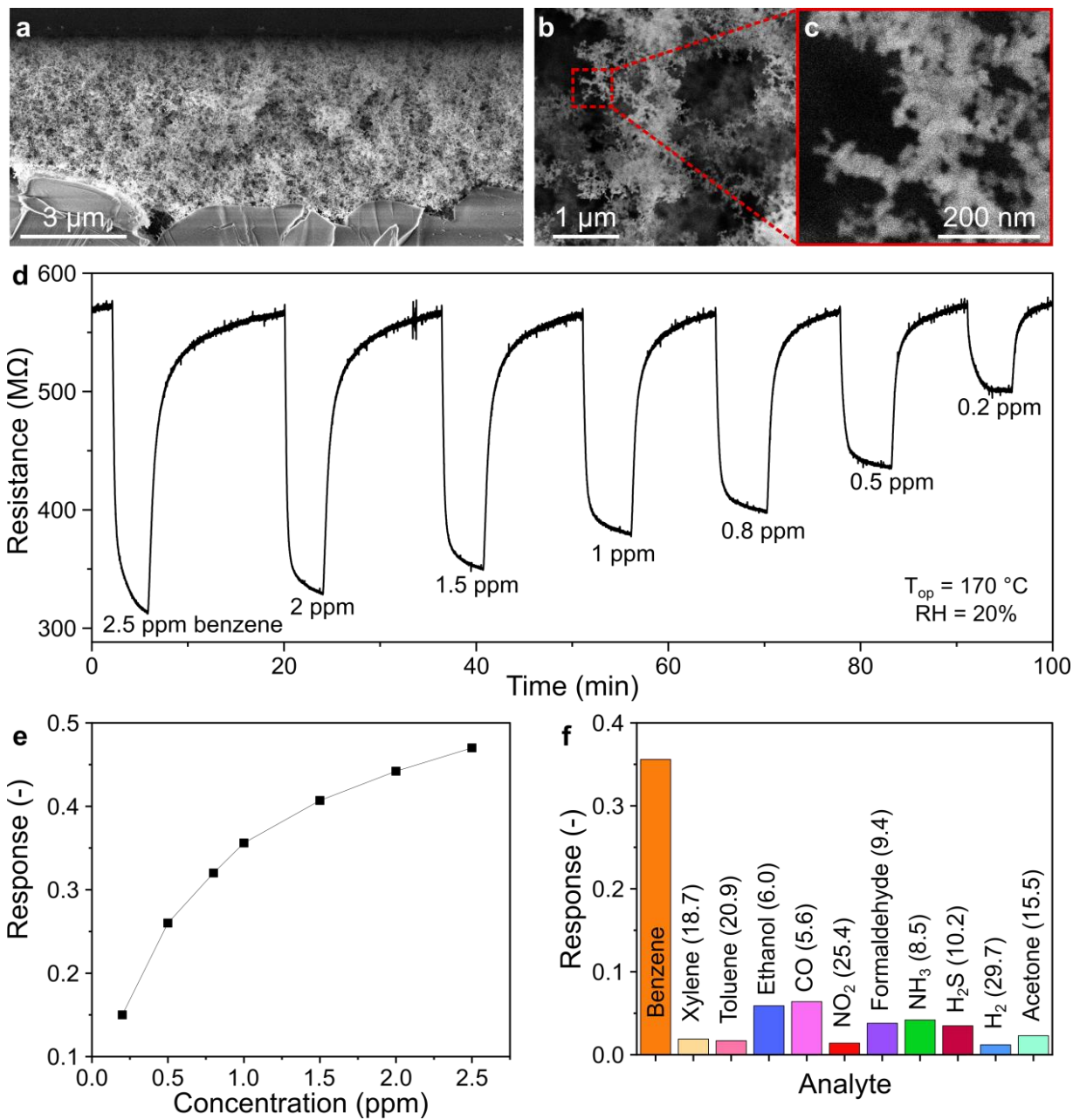

Fig. 5 (a) Cross-section and (b) top-view SEM of a porous $K_2W_7O_{22}$ porous film (K/W = 0.29) with a magnification in (c). Resistance (d) and corresponding responses (e) of a $K_2W_7O_{22}$ film operated at 170 °C during exposure to various concentrations of benzene at 20% RH in air. (f) Response of a porous $K_2W_7O_{22}$ film to different analytes at 1 ppm. Benzene selectivity ($S_{ben}/S_{conf}$) is shown in brackets.

# Supporting Information

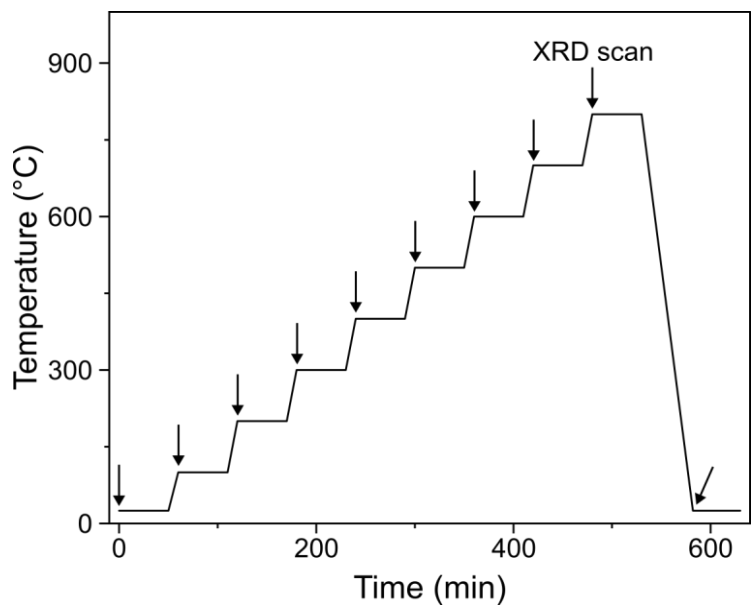

Fig. S1 Temperature profile of powder XRD measurement at elevated temperatures. Each arrow indicates the beginning of a scan.

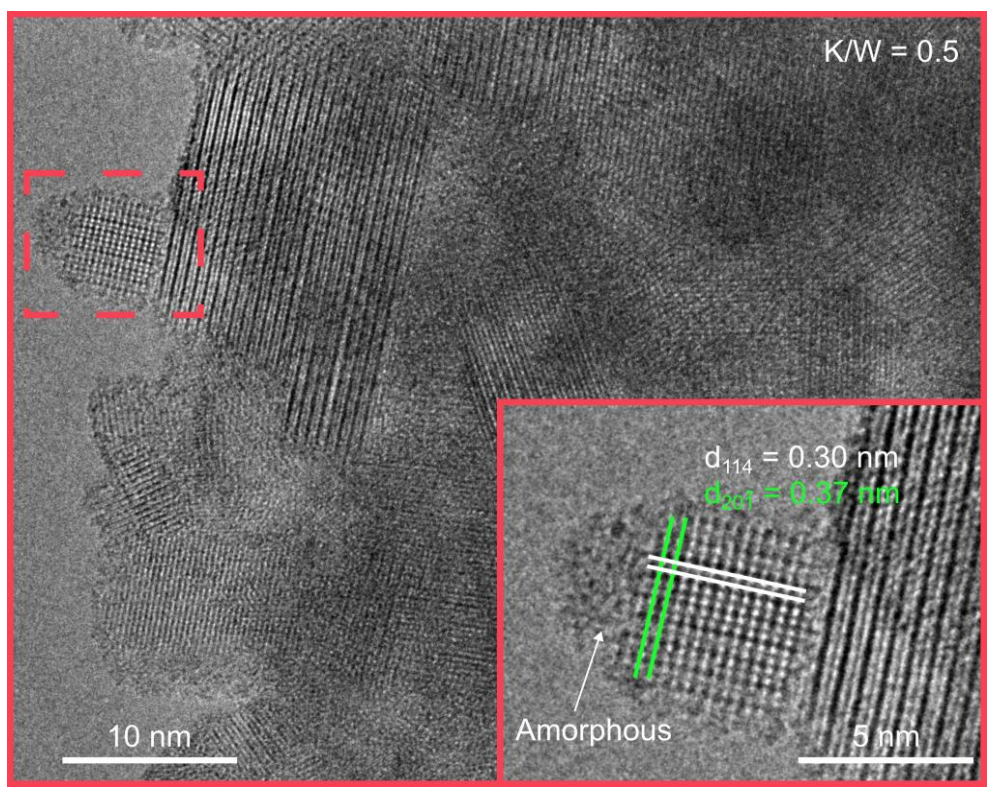

Fig. S2 HRTEM image of nanoparticles for K/W = 0.5. The inset shows a magnification of the indicated area revealing particles with partially crystalline and amorphous zones.

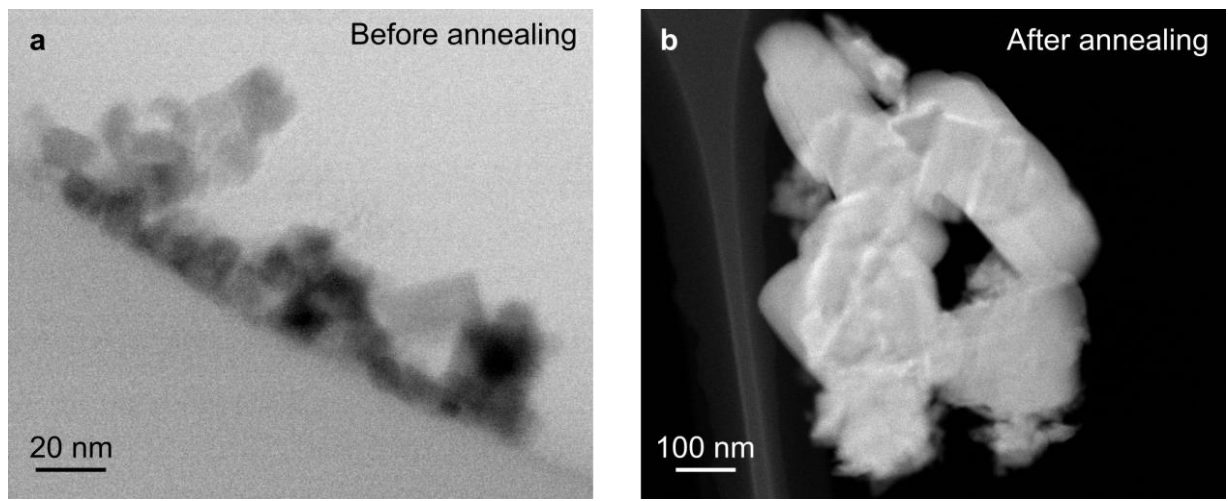

Fig. S3 TEM images of K$_2$W$_7$O$_{22}$ particles (a) before and (b) after annealing in ambient air until 800 °C following the temperature profile in Figure S1. Please note the different scale bars of the two images.

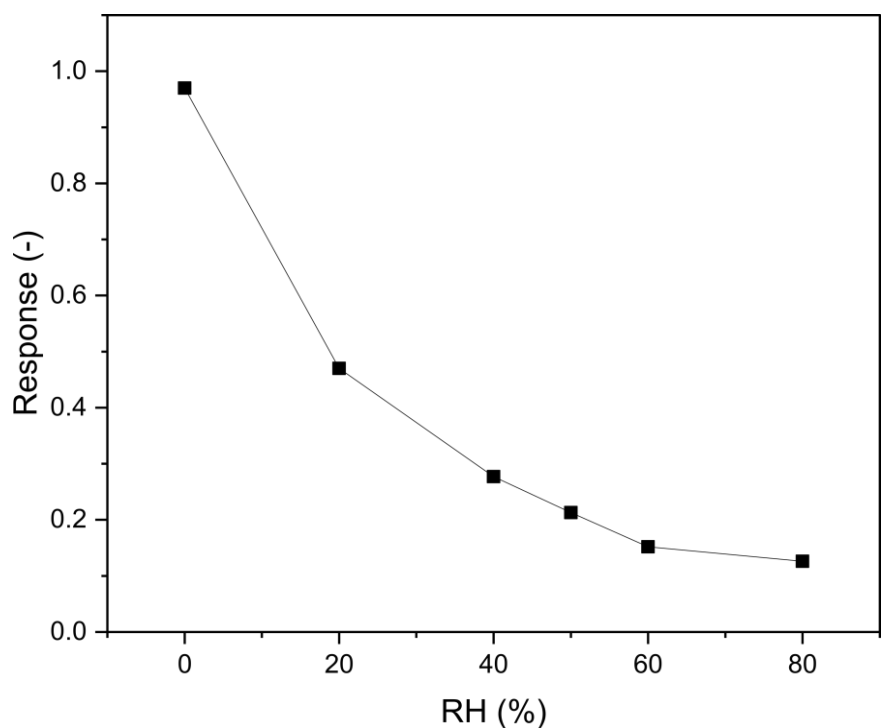

Fig. S4 Response to 1 ppm of benzene as a function of relative humidity (RH) for a K$_2$W$_7$O$_{22}$ (K/W = 0.29) porous sensing film operated at 170 °C in air.